\documentclass[12pt]{article}%
\usepackage{amsfonts}
\usepackage{cite}
\usepackage{amsmath}
\usepackage{amssymb}
\usepackage{graphicx}%
\providecommand{\U}[1]{\protect\rule{.1in}{.1in}}
\makeatletter
\@addtoreset{equation}{section}
\makeatother

\begin{document}
\begin{titlepage}
\begin{center}
\renewcommand{\thefootnote}{\fnsymbol{footnote}}
{\Large{\bf Euclidean Supergravity and Multi-Centered Solutions}}
\vskip 1.3cm
W. A. Sabra
\vskip 1cm
{\small{\it
Centre for Advanced Mathematical Sciences and Physics Department\\
American University of Beirut\\ Lebanon  \\}}
\end{center}
\bigskip
\begin{center}
{\bf Abstract}
\end{center}
In ungauged supergravity theories, the no-force condition for BPS states implies the existence of stable static
multi-centered solutions. The first solutions to Einstein-Maxwell theory with a positive cosmological constant
describing an arbitrary number of charged black holes were found by Kastor and Traschen.
Generalisations to five and higher dimensional theories were obtained by London.
Multi-centered solutions in gauged supergravity, even with time-dependence allowed,
have yet to be constructed. In this letter we construct supersymmetry-preserving multi-centered solutions for
the case of $D=5$, $N=2$ Euclidean gauged supergravity coupled to an arbitrary number of vector multiplets. Higher dimensional Einstein-Maxwell multi-centered solutions are also presented.
\end{titlepage}\ \ 

\section{Introduction}

In Newtonian gravity, one can obtain a system of point particles, each having
a charge equal to its mass, in static equilibrium by balancing the mutual
attractive gravitational and repulsive electrostatic forces. In Einstein
theory of general relativity and quite surprisingly the analogue situation
first appeared in the early work of Weyl, Majumdar and Papapetrou \cite{wmp}.
The Majumdar-Papapetrou (MP) solutions were found to describe a system of
multi-centered extremal Reissner-Nordstr\"{o}m black holes\ in thermal and
mechanical equilibrium \cite{hh}. The MP metrics are the static limits of the
Israel-Wilson-Perj\'{e}s (IWP) solutions \cite{iwp}. If one considers
Einstein-Maxwell theory as the bosonic sector of the theory of $N=2,$ $D=4$
supergravity, then the MP metrics turn out to be solutions admitting half of
the supersymmetry \cite{cmt}. A systematic classification by Tod \cite{Tod}
also demonstrated that the IWP\ metrics are the unique solutions with
time-like Killing vector admitting supercovariantly constant spinors.
Analogues of the MP solutions were also found for black holes with a dilaton
field in \cite{dmp}. In \cite{bls} general half-supersymmetric solutions,
which can be considered as generalisation of the MP and IWP metrics, to the
theories $N=2,$ $D=4$ supergravity with vector multiplets were found.
BPS\ solutions in five-dimensional Einstein-Maxwell theory were considered in
\cite{gkltt}. The metric in this case is of the Tanghelini form \cite{frt}.
Moreover, electric and magnetic BPS solutions breaking half of the
supersymmetry in ungauged five-dimensional supergravity coupled to vector
multiplets were constructed in \cite{sabra}.

The first multi-centered solutions asymptotic to de Sitter space were obtained
in four dimensions in \cite{kt}. These are non-static solutions to the
Einstein-Maxwell equations in the presence of a positive cosmological
constant. They describe an arbitrary number of charged black holes in motion
due to the positive cosmological constant. As should be expected, these
solutions reduce to MP solutions in the limit of vanishing cosmological
constant. Multi-centered solutions to $d$-dimensional Einstein-Maxwell theory
with a positive cosmological constant were given in \cite{london}. In the
five-dimensional case and with an imaginary coupling $g$ (with the
cosmological constant being proportional to $-g^{2})$, this theory may be
viewed as the bosonic sector of pure de Sitter $D=5$, $N=2$ supergravity.
Within this fake supergravity framework, it was shown in \cite{london} that
the multi-centered solutions preserve some supersymmetry through the explicit
construction of the corresponding Killing spinors. Multi-centered solutions to
$D=5$, $N=2$ gauged supergravity coupled to an arbitrary number of vector
multiplets were considered in \cite{jwks, kc}. In the fake de Sitter
supergravity case, one obtains rotating multi-centered solutions. However, in
the standard Anti-de Sitter cases, the multi-centered solutions have a complex
space-time metric.

In this letter, we will be mainly concerned with the gauged version of the
five-dimensional Euclidean supergravity theory which was recently constructed
in \cite{e5}. It will be demonstrated through the analysis of the Killing
spinor equations that this theory admits multi-centered solutions with real
space-time metric and real fields unlike in the Lorentzian theory where the
solution is complex. The results obtained may be useful as many investigations
of the AdS/CFT conjecture \cite{ads} have in fact been performed in Euclidean space.

We organise our work as follows. In section two we briefly present the
Euclidean five-dimensional theory and its special geometry structure relevant
for our subsequent analysis. Section three contains the analysis and
construction of the multi-centered solution for these theories. In the last
section we present multi-centered solutions to $d$-dimensional
Einstein-Maxwell theories and end with a summary.

\section{Euclidean $D=5$, $N=2$ Gauged Supergravity}

The theory of $D=5$, $N=2$ Euclidean supergravity coupled to vector multiplets
was recently constructed in \cite{e5}. The new Euclidean theory has the same
bosonic fields content of the Lorentzian theory with scalar fields
parametrizing a projective special real target manifold \cite{Gunaydin}. The
Lagrangian of the Euclidean theory differs from the Lorentzian one in that the
terms of the Euclidean gauge fields appear with the opposite sign. Upon
dimensional reduction on a circle, the Euclidean $D=5 $, $N=2$ bosonic
Lagrangian produces the bosonic Lagrangian of $D=4$, $N=2$ Euclidean
supergravity with the `wrong' sign in front of the gauge terms. However, in
four dimensions with Euclidean signature, theories with different signs in
front of their gauge kinetic terms can be mapped to one another by a duality
transformation \cite{e5, cosinst}.

In the five-dimensional Euclidean supergravity cases, unlike in four
dimensions, only one sign is allowed in front of the gauge field terms in the Lagrangian.

As the scalar fields structure in the Euclidean theory is unaltered, a scalar
potential $V$ can be added to the theory resulting in the gauged Euclidean
$N=2$, $D=5$ supergravity coupled to an arbitrary number $n$ of abelian vector
supermultiplets. The action is given by
\begin{align}
S  &  =\frac{1}{16}\int\left(  R-2g^{2}V\right)  \ast1-G_{IJ}\left(
dX^{I}\wedge\ast dX^{J}-F^{I}\wedge\ast F^{J}\right) \nonumber\\
&  -\frac{C_{IJK}}{6}F^{I}\wedge F^{J}\wedge A^{K}, \label{lage}%
\end{align}
where $I,$ $J$ take values $1,...,n$, $R$ is the scalar curvature,
$F^{I}=dA^{I}$ denote the abelian field-strengths two-forms. The constants
$C_{IJK}$ are symmetric in all indices and the coupling matrix $G_{IJ}$ is
invertible and remains the same as in the Lorentzian case. The fields $X^{I}$
are functions of $(n-1)$ unconstrained scalars $\phi^{i}.$ Some useful
relations which from very special geometry which will be used in our analysis
are
\begin{align}
G_{IJ}  &  =\frac{1}{2}\left(  9X_{I}X_{J}-C_{IJK}X^{K}\right)  ,\nonumber\\
\frac{1}{6}C_{IJK}X^{I}X^{J}X^{K}  &  =X^{I}X_{I}=1,\qquad\nonumber\\
dX_{I}  &  =-\frac{2}{3}G_{IJ}dX^{J},\hbox{ \ \ \ \ \ \ }X_{I}=\frac{2}%
{3}G_{IJ}X^{J},\qquad\nonumber\\
X_{I}dX^{I}  &  =X^{I}dX_{I}=0.
\end{align}
\ The scalar potential of the theory is given by%
\begin{equation}
V=9V_{I}V_{J}\left(  X^{I}X^{J}-\frac{1}{2}G^{IJ}\right)  ,
\end{equation}
where the $V_{I}$ are constants. The Killing spinor equations are given by
\footnote{Our conventions are as follows: We use the metric $\eta
^{ab}=(+,+,+,+,+)$ and Clifford algebra $\{{\gamma^{a},\gamma^{b}}%
\}=2\eta^{ab}$. The covariant derivative on spinors is $\nabla_{\mu}%
=\partial_{\mu}+{\frac{1}{4}}\omega_{\mu ab}\gamma^{ab}$ where $\omega_{\mu
ab}$ is the spin connection. Finally, antisymmetrization is with weight one,
so $\gamma^{{a}_{1}{a}_{2}\cdots{a_{n}}}={\frac{1}{n!}}\gamma^{\lbrack{a_{1}}%
}\gamma^{{a_{2}}}\cdots\gamma^{{a_{n}}]}$.} \cite{e5}
\begin{align}
\left[  \nabla_{\mu}+\frac{3}{2}gV_{I}A_{\mu}^{I}-\frac{1}{8}X_{I}(\gamma
_{\mu}{}^{\nu\rho}-4\delta_{\mu}^{\nu}\gamma^{\rho})F_{\nu\rho}^{I}+\frac
{1}{2}gX^{I}V_{I}\gamma_{\mu}\right]  \epsilon &  =0,\label{kil1}\\
\left(  3\partial_{\mu}X_{I}\gamma^{\mu}-G_{IJ}F_{\mu\nu}^{J}\gamma^{\mu\nu
}+6gV_{I}\right)  \partial_{i}X^{I}\epsilon &  =0, \label{kil2}%
\end{align}
where $\partial_{i}$ denotes differentiation with respect to the scalars
$\phi^{i}.$

\section{Supersymmetric Multi-centered Solutions}

In this section we construct multi-centered solutions to Euclidean theory
described by (\ref{lage}). Before proceeding to construct the solutions of the
gauged theory, we start with the analysis of the solutions of the ungauged
theory, i. e, for the cases when $g=0$. As in \cite{sabra}, we start with the
following metric ansatz
\begin{equation}
ds^{2}=e^{-4U}(d\tau+w)^{2}+e^{2U}ds_{4}^{2},
\end{equation}
with $U$ and $w=w_{m}dx^{m}$ independent of the $\tau$ coordinate. The
four-dimensional base space described by $ds_{4}^{2}$ is flat Euclidean with
coordinates $x^{m}$. The spin-connections components can be extracted from the
vanishing of torsion conditions
\begin{equation}
d\mathbf{e}^{0}+\omega^{0a}\wedge\mathbf{e}^{a}%
=0,\hbox{ \ \ \ \ \ \ }d\mathbf{e}^{a}+\omega^{ab}\wedge\mathbf{e}^{b}%
-\omega^{0a}\wedge\mathbf{e}^{0}=0,
\end{equation}
where%

\begin{equation}
\mathbf{e}^{0}=e^{-2U}\left(  d\tau+w\right)  ,\hbox{ \ \ \ }\mathbf{e}%
^{a}=e^{U}\delta_{m}^{a}dx^{m},
\end{equation}
and are given by%

\begin{align}
\omega_{\tau}^{0a}  &  =-2e^{-3U}\delta^{am}\partial_{m}U,\nonumber\\
\omega_{n}^{0a}  &  =-e^{-3U}\delta^{am}\left(  \frac{1}{2}w_{nm}%
+2w_{n}\partial_{m}U\right)  ,\nonumber\\
\omega_{\tau}^{ab}  &  =\frac{1}{2}e^{-6U}\delta^{nb}\delta^{ma}%
w_{nm},\nonumber\\
\omega_{n}^{ab}  &  =(\delta^{mb}\delta_{n}^{a}-\delta^{ma}\delta_{n}%
^{b})\partial_{m}U+\frac{1}{2}e^{-6U}w_{n}\delta^{pb}\delta^{ma}w_{pm},
\end{align}
where $w_{nm}=(\partial_{n}w_{m}-\partial_{m}w_{n}).$ Plugging into the
Killing spinor equation (\ref{kil1}) for $g=0,$ requiring that the Killing
spinor satisfies the projection condition $\gamma^{0}\epsilon=\epsilon,$ and
making use of the identity $\gamma_{abc}{}=-\varepsilon_{abcd}\gamma^{d}%
\gamma_{0}$ and special geometry relations we obtain the conditions%

\begin{align}
\partial_{\tau}\epsilon &  =0,\nonumber\\
F_{\tau m}^{I}  &  =\partial_{m}\left(  X^{I}e^{-2U}\right)  ,\nonumber\\
F_{mn}^{I}  &  =\partial_{n}\left(  e^{-2U}X^{I}w_{m}\right)  -\partial
_{m}\left(  e^{-2U}X^{I}w_{n}\right)  , \label{gfu}%
\end{align}
and
\begin{align}
\left(  \partial_{n}+\partial_{n}U\right)  \epsilon &  =0,\label{dif}\\
\gamma^{b}\left(  w_{ab}+\frac{1}{2}\varepsilon_{abcd}w^{cd}\right)
\epsilon{}  &  =0.
\end{align}
Therefore we get%

\begin{equation}
dw=-\ast dw, \label{duality}%
\end{equation}
thus implying that the two-form $\phi=dw$ satisfies%

\begin{equation}
d\phi=d\ast\phi=0,
\end{equation}
and therefore is a harmonic two-form. Equation (\ref{dif}) implies that the
Killing spinor is given by
\begin{equation}
\epsilon=e^{-U}\epsilon_{0},\hbox{ \ \ \ \ }\gamma_{0}\epsilon_{0}%
=\epsilon_{0},
\end{equation}
where $\epsilon_{0}$ is a constant spinor. It can then be shown that the
equation (\ref{kil2}) for $g=0$ is satisfied for scalars independent of $\tau
$. The Bianchi identities for our solution are identically satisfied. Using
(\ref{gfu}) and (\ref{duality}) we find that the Maxwell equations
\begin{equation}
d\left(  G_{IJ}\ast F^{J}\right)  =\frac{1}{4}C_{IJK}F^{J}\wedge F^{K}%
\end{equation}
are satisfied provided%

\begin{equation}
X_{I}=\frac{1}{3}e^{-2U}H_{I}, \label{flow}%
\end{equation}
where $H_{I}$ are a set of harmonic functions,%

\begin{equation}
H_{I}=h_{I}+%
%TCIMACRO{\dsum \limits_{j=1}^{N}}%
%BeginExpansion
{\displaystyle\sum\limits_{j=1}^{N}}
%EndExpansion
\frac{q_{Ij}}{|\vec{x}-\vec{x}_{j}|^{2}}.
\end{equation}
Here $h_{I}$ are related to scalar values at infinity and $q_{Ij}$ are
electric charges. As in \cite{sabra}, we define the rescaled coordinates%

\begin{equation}
Y_{I}=e^{2U}X_{I},\hbox{ \ \ \ }Y^{I}=e^{U}X^{I},
\end{equation}
then the solution for $U$ is given by%

\begin{equation}
e^{3U}=\frac{1}{6}C_{IJK}Y^{I}Y^{J}Y^{K},\hbox{ \ \ \ \ \ }\frac{1}{2}%
C_{IJK}Y^{J}Y^{K}=H_{I}. \label{sol}%
\end{equation}
To get explicit solutions for a given model one needs to solve for the
equations (\ref{sol}) which depend on the intersection numbers $C_{IJK}$.
However one can get a closed general solution when the scalar fields take
values in a symmetric space where one have the useful condition
\cite{Gunaydin}
\begin{equation}
C_{IJK}C_{J^{\prime}(LM}C_{PQ)K^{\prime}}\delta^{JJ^{\prime}}\delta
^{KK^{\prime}}={\frac{4}{3}}\delta_{I(L}C_{MPQ)}.
\end{equation}
In this case we have the identity
\begin{equation}
X^{I}=\frac{9}{2}C^{IJK}X_{J}X_{K}\hbox{ },
\end{equation}
where $C^{IJK}=\delta^{II^{\prime}}\delta^{JJ^{\prime}}\delta^{KK^{\prime}%
}C_{I^{\prime}J^{\prime}K^{\prime}}$, then the solution (\ref{sol}) implies%

\begin{equation}
e^{6U}=\frac{1}{6}C^{IJK}H_{I}H_{J}H_{K}.
\end{equation}
We now move on to construct multi-centered solutions for Euclidean
five-dimensional supergravity theories with non-trivial gauge and scalar
fields. It will be shown that in the Euclidean case multi-centred solutions
with real space-time metric do exist. Motivated by the results of \cite{jwks},
we take as an ansatz for our solution the metric
\begin{equation}
ds^{2}=e^{-4U}(d\tau+e^{2g\tau}w)^{2}+e^{-2g\tau}e^{2U}ds_{4}^{2},
\end{equation}
where $U=U(x,\tau)$, $w=w_{m}(x)dx^{m}$ depends on the base coordinates only.
The spin-connections components can be extracted from the vanishing of torsion
conditions and are given by%

\begin{align}
\omega_{\tau}^{0a}  &  =e^{g\tau-U}\delta^{am}\left(  \partial_{m}\left(
e^{-2U}\right)  -\partial_{\tau}Q_{m}\right)  ,\nonumber\\
\omega_{n}^{0a}  &  =-e^{3U-g\tau}(\dot{U}-g)\delta_{n}^{a}-\frac{1}%
{2}e^{3\left(  g\tau-U\right)  }\delta^{am}w_{nm}\nonumber\\
&  +e^{g\tau+U}\delta^{am}Q_{n}\left(  \partial_{m}\left(  e^{-2U}\right)
-\partial_{\tau}Q_{m}\right)  ,\nonumber\\
\omega_{\tau}^{ab}  &  =\frac{1}{2}e^{4g\tau-6U}\delta^{nb}\delta^{ma}%
w_{nm},\nonumber\\
\omega_{n}^{ab}  &  =(\delta^{mb}\delta_{n}^{a}-\delta^{ma}\delta_{n}%
^{b})\left(  \partial_{m}U+\frac{1}{2}e^{2U}\partial_{\tau}Q_{m}\right)
+\frac{1}{2}e^{4\left(  g\tau-U\right)  }Q_{n}\delta^{pb}\delta^{ma}w_{pm},
\end{align}
where $Q_{m}=e^{2\left(  g\tau-U\right)  }w_{m}.$ Plugging this into the
Killing spinor equation (\ref{kil1}) and as in the ungauged case we require
that the Killing spinor satisfies $\gamma^{0}\epsilon=\epsilon,$ we then get
from the $\tau$-component the conditions%

\begin{equation}
\left(  \partial_{\tau}-gV_{I}e^{-2U}X^{I}\right)  \epsilon=0, \label{tau}%
\end{equation}
and%
\begin{align}
X_{I}F_{\tau m}^{I}  &  =\partial_{m}\left(  e^{-2U}\right)  -\partial_{\tau
}Q_{m},\nonumber\\
X_{I}F_{mn}^{I}  &  =\partial_{n}Q_{m}-\partial_{m}Q_{n}.
\end{align}
Using the special geometry relations $X_{I}dX^{I}=0$ and $X^{I}X_{I}=1$, we
can write%

\begin{align}
F_{\tau m}^{I}  &  =\partial_{m}\left(  X^{I}e^{-2U}\right)  -\partial_{\tau
}\left(  X^{I}Q_{m}\right)  ,\nonumber\\
F_{mn}^{I}  &  =\partial_{n}\left(  X^{I}Q_{m}\right)  -\partial_{m}\left(
X^{I}Q_{n}\right)  ,
\end{align}
and therefore the gauge field can be given by%

\begin{equation}
A_{m}^{I}=-X^{I}Q_{m},\hbox{ \ \ }A_{\tau}^{I}=-e^{-2U}X^{I}.
\end{equation}
The rest of the components of (\ref{kil1}) give, in addition to the conditions
obtained in the ungauged case (\ref{dif}) and (\ref{duality}), the following condition%

\begin{equation}
gX^{I}V_{I}+e^{2U}(-g+\dot{U})=0. \label{dcon}%
\end{equation}
The equations (\ref{tau}) and (\ref{dcon}) then imply
\begin{equation}
\left(  \partial_{\tau}-g+\dot{U}\right)  \epsilon=0,
\end{equation}
which together with (\ref{dif}) imply that the Killing spinor equations are
solved by%

\begin{equation}
\epsilon=e^{g\tau-U}\epsilon_{0},\hbox{ \ \ \ }\gamma_{0}\epsilon_{0}%
=\epsilon_{0}.
\end{equation}
Turning to the second equation (\ref{kil2}) and substituting the equations
obtained so far, we obtain the condition%

\begin{equation}
\left(  3e^{2U}\partial_{\tau}X_{I}+6gV_{I}\right)  \partial_{i}X^{I}=0.
\label{gacon}%
\end{equation}
Using special geometry relations, the $\gamma^{a}$ and the $\gamma^{ab}$ terms
vanish identically. Note that $F^{I}=dA^{I},$ where the gauge fields one-forms
are given by%

\begin{equation}
A^{J}=-X^{J}\mathbf{e}^{0},
\end{equation}
then the Bianchi identities hold automatically.

After some analysis it can be shown that the condition (\ref{gacon}) and the
Maxwell's equations are satisfied for our solution provided that the scalars satisfy%

\begin{align}
e^{2U}X_{I}  &  =\frac{1}{3}H_{I},\nonumber\\
H_{I}(t,\vec{x})  &  =3V_{I}+e^{2g\tau}\sum_{j=1}^{N}{\frac{q_{I\,j}}{|\vec
{x}-\vec{x}_{j}|^{2}}}%
\end{align}
together with the condition%

\begin{equation}
d\ast_{4}w=0.
\end{equation}

This completes the construction of the multi-centered solutions to $D=5$,
$N=2$ gauged Euclidean supergravity theories.

\section{$d$-dimensional Solutions}

In this section we present general multi-centered solutions to $d$-dimensional
Einstein-Maxwell theory with a cosmological constant. We start first by
writing the $d$-dimensional de Sitter space-time metric in terms of the
so-called cosmological coordinates, this is given by
\begin{equation}
ds^{2}=-d\tau^{2}+e^{-2l\tau}ds_{(d-1)}^{2}, \label{met}%
\end{equation}
where $ds_{(d-1)}^{2}$ is the metric of $(d-1)$-dimensional flat Euclidean
space. The metric (\ref{met}) is a solution of $d$-dimensional Einstein
gravity with a positive cosmological constant with
\begin{equation}
R_{\mu\nu}=(d-1)l^{2}g_{\mu\nu}.
\end{equation}
Moreover, the metric
\begin{equation}
ds^{2}=d\tau^{2}+e^{-2l\tau}ds_{(d-1)}^{2},
\end{equation}
is a solution of $d$-dimensional Euclidean Einstein gravity with a negative
cosmological constant with
\begin{equation}
R_{\mu\nu}=-(d-1)l^{2}g_{\mu\nu}.
\end{equation}
We now consider a general $d$-dimensional Einstein-Maxwell with a cosmological
constant, with Lagrangian density%

\begin{equation}
e^{-1}\mathcal{L}_{d}=\!\!R+\eta F_{\mu\nu}F^{\mu\nu}-\Lambda.
\end{equation}
Here $\eta$ can be either $+1$ or $-1$. The Einstein gravitational equations
of motion are given by%
\begin{equation}
R_{\mu\nu}=-{2\eta}\left(  F_{\mu\lambda}F_{\nu}{}^{\lambda\,}-{\frac
{1}{2(d-2)}}g_{\mu\nu}F_{\rho\sigma}F^{\rho\sigma}\right)  +{\frac{g_{\mu\nu}%
}{d-2}\Lambda}. \label{m}%
\end{equation}
If one considers the solution%
\begin{equation}
ds^{2}=\frac{\eta}{H^{2}}d\tau^{2}+H^{2/(d-3)}e^{-2l\tau}ds_{d-1}^{2}%
\end{equation}
then it can be shown that this is a solution of (\ref{m}) provided that%

\begin{align}
{\Lambda}  &  =-\eta(d-1)(d-2)l^{2},\nonumber\\
F_{\tau i}{}^{\,}  &  =\sqrt{\frac{\left(  d-2\right)  }{2\left(  d-3\right)
}}\frac{\partial_{i}H}{H^{2}},\nonumber\\
H  &  =1+\sum_{j=1}^{d-1}{\frac{q_{\,j}}{|\vec{x}-\vec{x}_{j}|^{d-3}}%
}e^{(d-3)l\tau}.
\end{align}

Here $\partial_{i}$ represents differentiation with respect to the coordinates
of the base space described by $ds_{d-1}^{2}$. For $\eta=-1,$ we reproduce the
solutions of \cite{kt, london}, i. e., multi-centrered solutions with a
positive cosmological constant. For $\eta=1,$ we obtain new multi-centered
solutions for the Euclidean theory with a negative cosmological constant at
the expense of introducing a Lagrangian with the opposite sign of the gauge
terms. The systematic analysis of the Euclidean four-dimensional cases was
treated in \cite{cosinst} \footnote{Note that if we consider the metrics
\begin{equation}
ds^{2}=\frac{1}{H^{2}}d\tau^{2}+H^{2/(d-3)}e^{-2l\tau}ds_{d-1}^{2},
\end{equation}
and take $ds_{d-1}^{2}$ to be a flat metric with a Lorentzian signature, then
these are solutions with $H$ satisfying the wave-equation in ($d-1)$%
-dimensional Minkowski space. Four-dimensional solutions of this type in the
framework of ungauged $N=2,D=4$ supergravity coupled to vector multiplets were
considered in \cite{phantom}. These solutions deserve further analysis.} .

To summarise, we have constructed multi-centered solutions for the gauged
Euclidean supergravity with vector multiplets as well as for $d$-dimensional
Einstein-Maxwell theories. It must be noted that the cosmological
Kastor-Traschen solution was obtained as a class of pseudo-supersymmetric
solutions in the systematic analysis of minimal $N=2,$ $D=4$ de Sitter
supergravity \cite{4ds}. In five dimensions, the systematic analysis of
\cite{sys} revealed that solutions admitting Killing spinors in
five-dimensional ungauged and de Sitter supergravity have respectively a
hyper-K\"{a}hler and hyper-K\"{a}hler torsion (HKT) manifold as a four
dimensional space. The multi-centered solutions of \cite{london, jwks,kc} are
special cases. A systematic classification for the solutions of the gauged
Euclidean theories should be carried out and we hope to report on this in the
future. Finally, it remains an open question to construct true multi-centered
solutions of the standard Lorentzian gauged Anti-de Sitter supergravity theories.

\bigskip

\textbf{Acknowledgements} : \ The author would like to thank Jos\'{e}
Figueroa-O'Farrill for a useful discussion. This work is supported in part by
the National Science Foundation under grant number PHY-1620505

\end{document}